\documentclass[twocolumn,groupedaddress,amsmath,amssymb,aps,prl,reprint,floatfix]{revtex4-1}

\usepackage{graphicx} 
\usepackage{dcolumn} 
\usepackage{bm} 
\usepackage{verbatim} 
\usepackage{hyperref} 
\usepackage{color}
\usepackage{enumitem}
\usepackage{chemmacros} 
\usepackage{float}
\usepackage{siunitx}
\usepackage{ulem}

\begin{document}
  
\newcommand\magentasout{\bgroup\markoverwith{\textcolor{magenta}{\rule[0.5ex]{2pt}{0.4pt}}}\ULon}

\title{Domain-Wall Ferroelectric Polarons in a two-dimensional \\ Rotor Lattice Model}
\author{Florian Kluibenschedl}
\affiliation{Institute of Science and Technology Austria (ISTA), am Campus 1, 3400 Klosterneuburg, Austria}
\author{Georgios M. Koutentakis}
\affiliation{Institute of Science and Technology Austria (ISTA), am Campus 1, 3400 Klosterneuburg, Austria}
\author{Ragheed Alhyder}
\affiliation{Institute of Science and Technology Austria (ISTA), am Campus 1, 3400 Klosterneuburg, Austria}
\author{Mikhail Lemeshko}
\affiliation{Institute of Science and Technology Austria (ISTA), am Campus 1, 3400 Klosterneuburg, Austria}
\date{\today}

\begin{abstract}
We demonstrate the formation of ferroelectric domain-wall polarons in a minimal two-dimensional lattice model of electrons interacting with rotating dipoles. Along the domain-wall, the rotors polarize in opposite directions, causing the electron to localize along a particular lattice direction. The rotor-electron coupling is identified as the origin of a structural instability in the crystal that leads to the domain-wall formation via a symmetry-breaking process. Our results provide the first theoretical description of ferroelectric polarons, as discussed in the context of soft semiconductors.
\end{abstract}

\maketitle  

The properties of many systems are governed by the rotational degrees of freedom associated with dipolar rotors embedded in a lattice environment~\cite{Bakulin2015,deng2013probable,Yokoyama2011,DeVlugt2020,Liehm2012,Burchesky2021,Holland2023}. Example systems include dipoles trapped in optical lattices~\cite{Micheli2006,BARANOV2008,Schachenmayer2010}, dipoles pinned to organic substrates~\cite{Perera2012,Lee2015,Hamer2021} or dipoles in crystalline materials~\cite{Miyata2018,Wang2021,Wang2022}. In the latter, the rotating dipoles may be attributed to either molecules embedded in the lattice~\cite{Bakulin2015}, or to anharmonic and dynamically disordered polar phonon modes behaving like effective dipoles~\cite{deng2013probable,Zacharias2023_1,Zacharias2023_2,Schilcher2021,Melnikov2023}. An overarching objective in such works is studying the collective orientation of the dipoles, which is often induced by dipole-dipole interactions or external control fields~\cite{Abolins2018,Magann2019,Ma2020} and leads to phenomena such as ferro- or paraelectricity~\cite{Chandra2017,Akutagawa2009,Zhang2012,Sachdev2011,Eglitis_2014}.

The collective orientation of the dipoles can also be initiated by introducing a charged impurity into the lattice, \textit{akin} to the polaron concept in condensed matter physics~\cite{alexandrov2009advances,Franchini2021}. In this scenario, the impurity couples to the rotating dipoles and forms a quasiparticle, the so-called ferroelectric polaron~\cite{Miyata2017,Wang2021,Koutentakis2023}. In contrast to the well-established Holstein-, or Fr{\" o}hlich-polarons~\cite{Yam2020,Froehlich1954,Bona1999,alexandrov2009advances}, the excess charge carrier in rotor lattices is dressed by the rotational excitations of the dipoles instead of the lattice vibrations, which significantly changes the observed physics~\cite{Koutentakis2023}.   

Experimental and {\it ab-initio} quantum chemistry studies identified the presence of (effective) dipoles in exemplary soft and polar semiconductors such as Lead Halide Perovskites (LHPs)~\cite{Quarti2015,Zhu2015,Yaffe2017,Miyata2017,Miyata2018,Qian2023} and \ch{Bi2O2Se}~\cite{Wang2022}. Calculations show that a polaron forms around an excess charge-carrier and that the polaron wavefunction tends to localize along different lattice directions, forming a domain-wall shaped polaron~\cite{Ambrosio2018,Ambrosio2019,Wang2021}. The dipoles in this type of polaron are arranged in two antioriented ferroelectric domains, with the electron confined within the corresponding domain-wall. The domain-wall formation is argued to shield the charge carrier from scattering with defects, phonons and other charges~\cite{Miyata2018,Wang2021,Wang2022}. This is one proposed explanation for the remarkable optoelectronic properties of soft semiconductors, such as their long carrier lifetimes and diffusion lengths~\cite{Manser2016,Brenner2016,Zhang2016}.

It is therefore anticipated that domain-wall ferroelectric polarons emerge in generic rotor lattice systems coupled to a charged impurity. Nevertheless, the minimal interaction mechanisms required for their emergence and their associated physical properties remain to be established. 

\begin{figure}[h]
\centering
\includegraphics[scale=0.35]{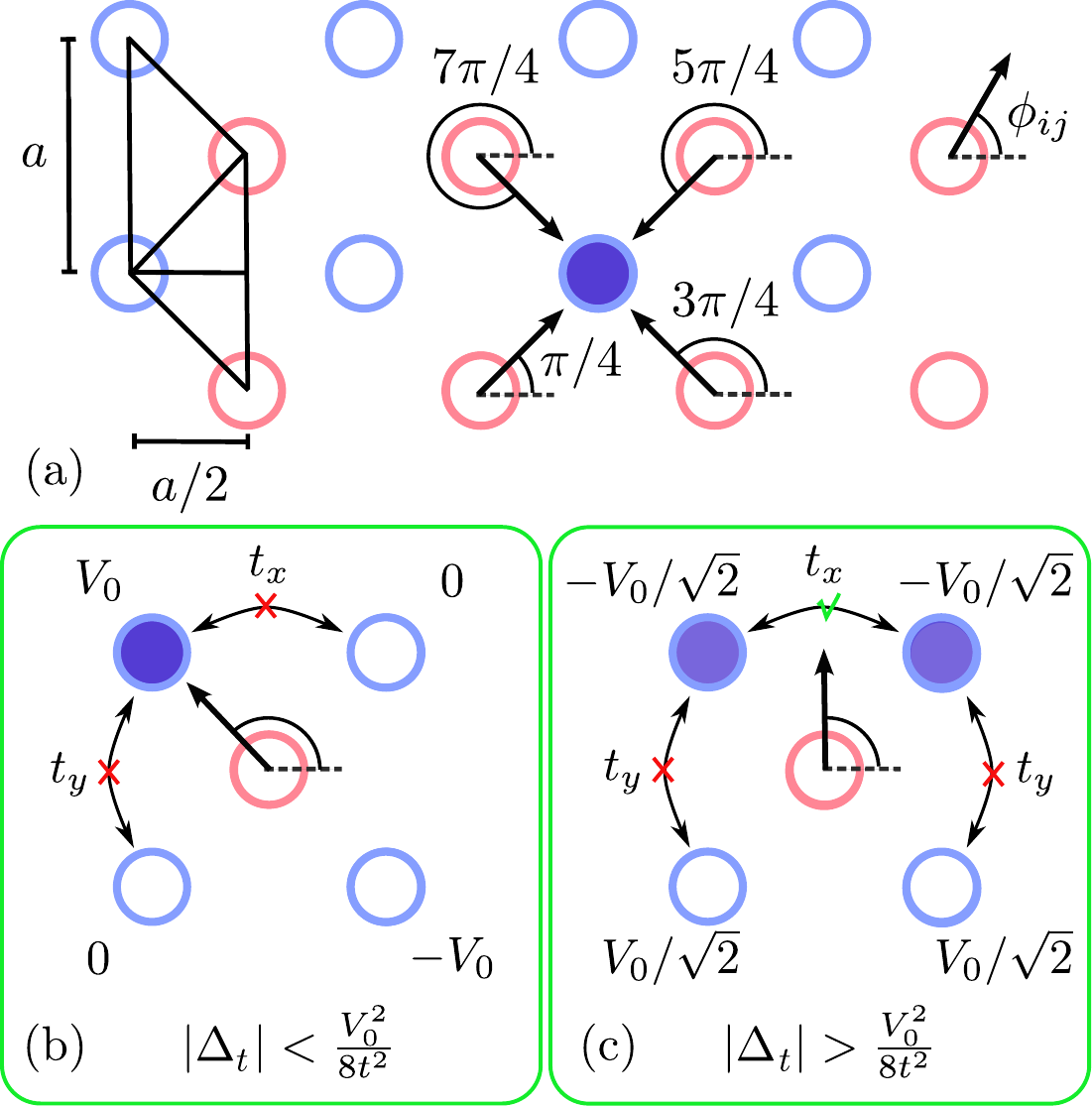}
\caption{(a) Geometry of the two-dimensional rotor-lattice model. The filled (empty) blue circles represent occupied (unoccupied) electron sites. The red circles indicate the positions of planar rotors. Rotor angles and displacement measures are provided. (b), (c) Schematic of the influence of two distinct rotor orientations on the hopping of the electron on an elementary four-site plaquette.
\label{fig:fig_model}}
\end{figure}
  
In this Letter, we demonstrate the emergence of ferroelectric domain-wall polarons in a minimal two-dimensional lattice model of dipolar rotors interacting with a mobile charge carrier, see Fig.~\ref{fig:fig_model}(a). In two dimensions, the orientation of the rotor controls the tunneling direction of the electron,~cf.~Fig.~\ref{fig:fig_model}(b) and~(c), leading to a richer interplay between electron localization and polaron transport than its one dimensional analogue~\cite{Koutentakis2023,Footnote}. Our variational phase diagram analysis reveals that domain-wall polarons occur for anisotropic systems when the rotor-electron attraction is of the same order as the electron kinetic energy. Moreover, the energetics of the system allow us to propose a pseudo-Jahn-Teller mechanism~\cite{englman1972jahn,Bersuker2006,Bersuker2016}, according to which these domains can arise spontaneously from an isotropic system through symmetry-breaking structural distortions. These distortions are associated with long-wavelength, tunneling modulating soft phonons, which are omnipresent in materials where ferroelectric domain-wall polarons are postulated to appear~\cite{Egger2016,Wang2021,Wang2022}. Beyond the regime of ferroelectric domain-wall polarons, we observe the formation of ferroelectric large and small polarons for weak and strong interactions, respectively. These polarons are characterized by isotropic rotor order and transport properties.

Our two-dimensional model is composed of two superimposed square lattices, pertaining to electron and rotors, see Fig.~\ref{fig:fig_model}(a). Both, the rotor and the electron lattice have a lattice constant $a$, and are shifted with respect to one another by $a/2$ in both the $x$ and $y$ lattice directions. Here, we consider a single electron band and planar rotors. The rotor orientation, $\phi_{j,l}$, with $(j,l)$ being the site index, is given by their angle measured from the horizontal axis. This rotor-lattice system is described by the Hamiltonian
\begin{equation} 
	\hat{H} = \hat{H}_{\text{hop}} + \hat{H}_{\text{rot}} + \hat{H}_{\text{int}}. \label{eq:hamiltonian}
\end{equation}
The first term, $\hat{H}_{\text{hop}} = -t_x \hat{\mathcal{T}}_x - t_y \hat{\mathcal{T}}_y$, describes the kinetic energy of the electron in the tight-binding approximation, moving along the $x$ and $y$ directions of the two-dimensional lattice. Here, we introduce the scaled kinetic energy operators,
\begin{equation}
\begin{split}
    \hat{\mathcal{T}}_x = \sum_{j,l} \hat{c}_{j,l+1}^{\dagger}\hat{c}_{j,l} + \text{h.c}, \, \hat{\mathcal{T}}_y = \sum_{j,l} \hat{c}_{j+1,l}^{\dagger}\hat{c}_{j,l} + \text{h.c.}, \label{eq:electron_kin_hamiltonian}
\end{split}
\end{equation}
where $\hat{c}_{i,j}$ ($\hat{c}_{i,j}^{\dagger}$) are the electron annihilation (creation) operators on the electron lattice site $\left(i,j\right)$ and $t_\mu$ are the tunneling integrals along the $\mu$-axis. The structure of the kinetic energy operators together with wavefunction normalization implies that for any single-electron state,~$\lvert\langle\hat{\mathcal{T}}_{\mu}\rangle\rvert \leq 2$. Furthermore, we employ periodic boundary conditions, i.e. $\hat{c}_{M_y+j,M_x+l} = \hat{c}_{j,l}$, where $M_\mu$ is the number of rotors along the $\mu \in \{ x, y \}$ direction, and the total rotor number is $M=M_x M_y$. The term $\hat{H}_{\text{rot}}$ is the total rotational energy,
\begin{equation}
	\hat{H}_{\text{rot}} = - B \sum_{j,l} \frac{\partial^2}{\partial \phi_{j,l}^2},\label{eq:rotor_kinetic_hamiltonian}
\end{equation}
with $B$ being the rotational constant, which is assumed  to be equal for all rotors. Finally, the third term in Eq.~\eqref{eq:hamiltonian} describes the electron-rotor interaction,    
\begin{equation}
\begin{split}
	&\hat{H}_{\text{int}} = V_0 \sum_{j,l} \hat{c}_{j,l}^{\dagger}\hat{c}_{j,l} \Bigl[\cos\left(\phi_{j-1,l} -\frac{\pi}{4}\right) + \cos\left(\phi_{j,l} + \frac{\pi}{4}\right) \\
    &+ \cos\left(\phi_{j-1,l-1} - \frac{3\pi}{4}\right) + \cos\left(\phi_{j,l-1} + \frac{3\pi}{4}\right)\Bigr], \label{eq:rotor_electron_interaction_hamiltonian}
\end{split}
\raisetag{0pt}
\end{equation}   
with coupling constant $V_0$. The angles are chosen so that the electron-rotor attraction is maximized when the respective rotor points towards the electron, see Fig.~\ref{fig:fig_model}(a). Rotor-rotor interactions are not considered, to focus on the dipolar order emerging from the rotor-electron coupling alone. In addition, such interactions are considered insignificant in certain setups postulated to feature ferroelectric polarons due to screening effects~\cite{Zhu2016}. In the present work, we restrict ourselves to $V_0 > 0$, which describes the interaction of dipolar rotors with electrons. Note that the behavior presented here will not change in the case of holes, $V_0 < 0$, since the single-rotor wavefunctions will be just inverted. We fix the units by setting~$t = t_x + t_y = 1$, which implies that the model is governed by the ratios $B/t$, $V_0/t$ and the asymmetry parameter~$\Delta_t = \left(t_x-t_y\right)/t$. 

In the case that $B/t\gtrsim 1$, long-range electron-rotor correlations are suppressed and the physics is dominated by the elementary excitations of the rotors adjacent to the electron~\cite{Koutentakis2023}, preventing the formation of intrinsically long-range domain-walls. However, in the case where $B/t\ll 1$, the interplay between the rotor-electron attraction, promoting rotor polarization, and the electron kinetic energy, being hindered by variations of the rotor polarization, can lead to large polaron formation. This regime is also relevant experimentally, since density functional theory (DFT,~\cite{Hohenberg_1964,Kohn_1965,Jones_2015}) calculations for LHPs showed that $t,V_0 \approx 0.1-\qty{1}{eV}$, while the molecular dipoles rotate much slower, because $B\approx\qty{1}{meV}$~\cite{Fabini2017,Kang2017}. The separation of rotational and vibrational time-scales allows the parametric treatment of electron-phonon interactions by renormalizing $t_{\mu}$~\cite{Footnote}. Hence, we focus on the regime~$B/t\ll 1, V_0/t\approx 1$.

In the limiting case, $B\rightarrow 0$, the rotors are classical, parameterized only by their orientation $\phi_{j,l}=\phi_{j,l}^{\text{cl}}$. The rotors are fully polarized and the electron localizes in the effective potential provided by the interaction term $H_{\text{int}}$ due to the orientation of the rotors. The way the rotor orientation affects the localization basin of the electron can be understood by considering a rotor surrounded by four electron sites. If a rotor is aligned towards a site, see Fig.~\ref{fig:fig_model}(b), it provides a large potential energy benefit or deficit for the sites it points towards or away from, respectively. This large potential variation suppresses the electron tunneling, resulting in a kinetic energy penalty. In contrast, if a rotor points between two electron sites, see Fig.~\ref{fig:fig_model}(c), it allows for tunneling between these two sites but prohibits tunneling in the perpendicular direction. Therefore, polarized rotors act like traffic control for the electron, either localizing it or allowing it to be transported along one lattice direction. The energetically preferable configuration strongly depends on the anisotropy $\Delta_t$, with single-site localization preferable for lower anisotropies and strong interactions $V_0 \gg t$, Fig.~\ref{fig:fig_model}(b), and the directed tunneling being preferable for high anisotropies and weaker interactions, Fig.~\ref{fig:fig_model}(c). In the lattice case, there exists one such elementary plaquette for each rotor, and the adjacent rotors effectively interact since their plaquettes share an edge, leading to an emerging rotor order~\cite{Footnote}. Notice that excited states of the electron are typically separated by energies of the order of $\text{min}(t,V_0)$, while shifting the center of localization results in a degenerate state since $\hat{H}$ is translational invariant.

In the semiclassical regime, $B>0, B\ll t,V_0$, the rotors in the spatial region where the electron is localized fluctuate weakly around their respective equilibrium orientations. In contrast, the rotors far away from the electron become completely depolarized, as they only interact very weakly with it. Therefore, the rotor order does not change significantly from the classical case. However, as previously stated, the interaction between the rotors and their neighbors within the electron localization basin lifts the degeneracy of states with different localization centers, thereby enabling the transport of the localized electron state.  

To capture this behavior, we employ the mean-field product ansatz \cite{atkins2011molecular,Dirac1930,frenkel1934wave,Beck2000} $\lvert\psi_{1,1}\rangle = \lvert\psi_{\text{ele}}\rangle \otimes_{j,l} \lvert\varphi_{j,l}\rangle$, by which we evaluate the localized state centered at the coordinate origin, $\lvert\psi_{1,1}\rangle$, \textit{via} its expansion in the variational electron, $\lvert\psi_{\text{ele}}\rangle$, and single-rotor states $\lvert\varphi_{j,l}\rangle$. Subsequently, to approximate the total state of the system, we diagonalize the full Hamiltonian in the basis provided by $\lvert\psi_{j,l}\rangle = \hat{T}_x^{l-1}\hat{T}_y^{j-1}\lvert\psi_{1,1}\rangle$, where $\hat{T}_{\mu}$ is the translation operator in the $\mu$ lattice direction, e.g. $\hat{T}_{x}\phi_{i,j} = \phi_{i+1,j}, \hat{T}_{x}\hat{c}_{i,j}^{\left(\dagger\right)} = \hat{c}_{i+1,j}^{\left(\dagger\right)}$. The above procedure is variational in nature and provides an adequate approximation in the semiclassical regime, $B \ll t, V_0$, where emergent quantum rotor-rotor correlations and electron excitations can be neglected.

As witnesses of the interplay between rotor-electron attraction and electron kinetic energy, we will use $\hat{\mathcal{T}_{\mu}}$ from Eq.~\eqref{eq:electron_kin_hamiltonian}, and the polarization vector,
\begin{equation}
     \hat{\boldsymbol{\mathcal{P}}}^{jl}_{\text{rot}} = \sum_{j',l'} \hat{c}_{j,l}^{\dagger}\hat{c}_{j,l} \left( \cos \phi_{j+j', l+l'} \boldsymbol{e}_x + \sin \phi_{j+j', l+l'} \boldsymbol{e}_y \right),
\end{equation}
which indicates the orientation of the rotor in the $(j,l)$-th rotor position relative to the electron. The localization characteristics of the electron wavefunction can be inferred from the distribution of the electron occupation amplitude $\mathcal{N}_{\text{ele}}^{jl} = \langle\psi\rvert\hat{c}_{jl}^{\dagger}\hat{c}_{jl}\lvert\psi\rangle$.

\begin{figure}[h]
\centering
\includegraphics[width=\linewidth]{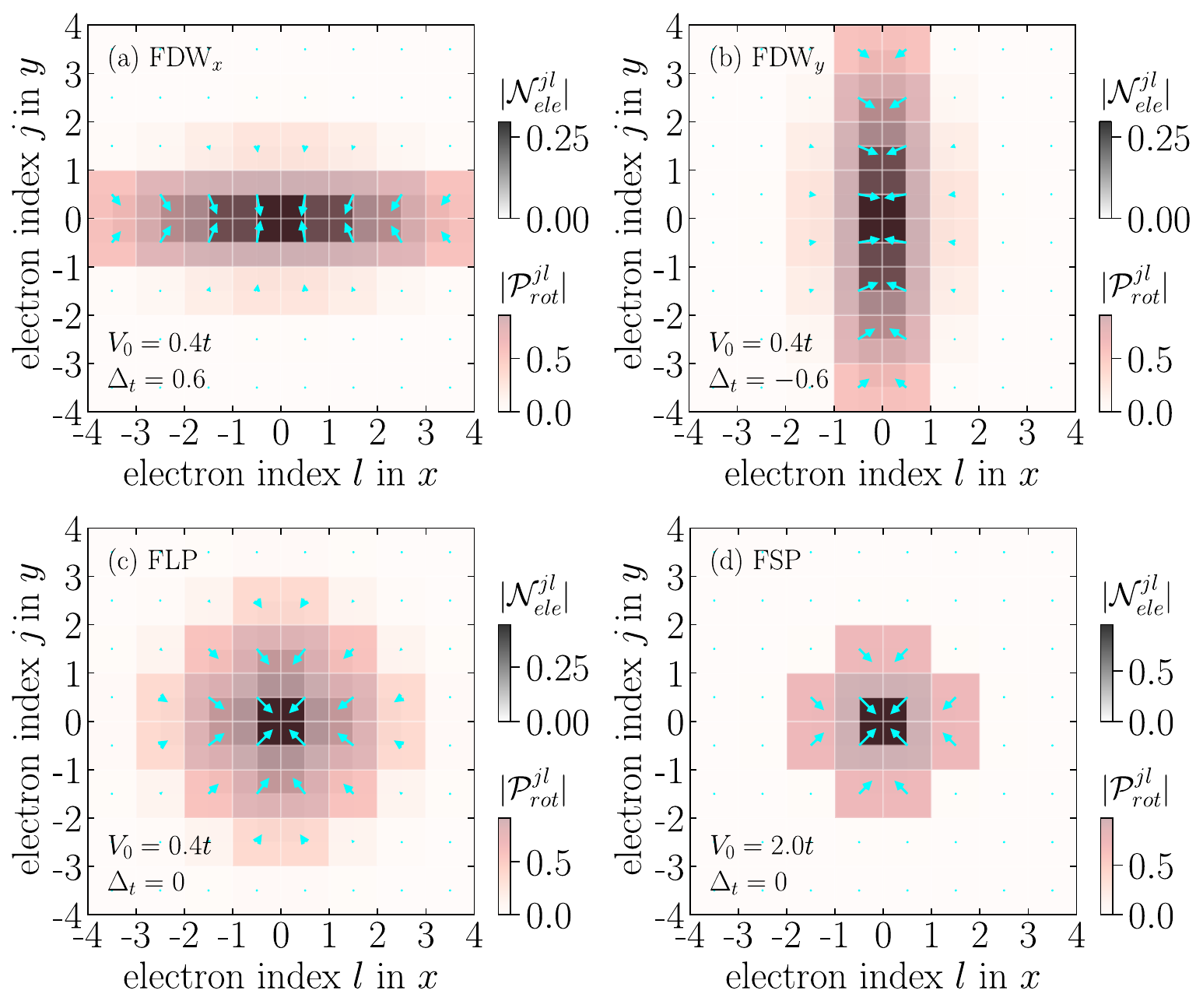}
\vskip -0.3cm
\caption{Magnitude of the polarization vector, $\boldsymbol{\mathcal{P}}^{jl}_{\text{rot}}$ (red), and the electron occupation amplitude $\mathcal{N}_{\text{ele}}^{jl}$ (black), both associated with the variational ground state of the electron-rotor lattice. The blue arrows indicate the orientation of the rotors, $\boldsymbol{\mathcal{P}}^{jl}_{\text{rot}}$, to illustrate the corresponding dipole order. For varying $V_0/t$ and $\Delta_t$, four distinct configurations arise: domain-wall states FDW$_x$ and FDW$_y$ in (a) and (b), FLP in (c) and FSP in (d). In all cases $B = 10^{-2}t$. \label{fig:fig_pol_size}}
\end{figure}  

Our variational analysis reveals the presence of four classes of ground states for varying $V_0$ and $\Delta_t$, each characterized by a unique arrangement of the dipole moments and the electron localization. This distinction can be illustrated by examining $\mathcal{N}_{\text{ele}}^{jl}$ and the orientation of polarization vectors $\boldsymbol{\mathcal{P}}^{jl}_{\text{rot}} = \langle\psi\rvert\hat{\boldsymbol{\mathcal{P}}}^{jl}_{\text{rot}}\rvert\psi\rangle$, see Fig.~\ref{fig:fig_pol_size}.

Remarkably, for asymmetric lattices, $\Delta_t \neq 0$, we find that the ground state distribution of the $\boldsymbol{\mathcal{P}}^{jl}_{\text{rot}}$ has a domain-wall structure on the rotor-lattice with rotational~$C_2$ symmetry, see Fig.~\ref{fig:fig_pol_size}(a) and \ref{fig:fig_pol_size}(b). The rotors form two opposite and largely polarized ferroelectric domains with a lateral extension of 8--10 rotors, depending on $V_0/t$. In the transverse direction, the rotor polarization extends over 2--4 rotors. The electron wave function is confined to the domain-wall, as illustrated by the distribution of $\mathcal{N}_{\text{ele}}^{jl}$. This is a manifestation of the mechanism described in Fig.~\ref{fig:fig_model}(c), enabling the transport of the electron perpendicularly to the rotor polarization. These states correspond to ferroelectric domain-wall polarons \cite{Wang2022} and herewith we refer to them as Ferroelectric Domain-Walls (FDW$_{\mu}$). Their wavefunction anisotropy leads to an anisotropic contribution to its kinetic energy. In particular, for a FDW$_\mu$, $1<\mathcal{T}_{\mu}=\langle\psi\rvert\hat{\mathcal{T}}_{\mu}\rvert\psi\rangle \leq 2$, while~$\mathcal{T}_{\eta\neq\mu} < 1$.

In contrast, the distributions of $\mathcal{N}_{\text{ele}}^{jl}$ and $\boldsymbol{\mathcal{P}}^{jl}_{\text{rot}}$ are $C_4$ symmetric in the weak and strong coupling regimes for $\Delta_t = 0$, see Fig.~\ref{fig:fig_pol_size}(c) and \ref{fig:fig_pol_size}(d), where the $\boldsymbol{\mathcal{P}}^{jl}_{\text{rot}}$ vectors always point towards the electron, since the mechanism of Fig.~\ref{fig:fig_model}(b) is dominant. For weak couplings, the polarization cloud extends over 6--8 rotors in both lattice directions and we call the respective state Ferroelectric Large Polaron (FLP). In this state, the electron kinetic energy dominates and thus it is characterized by $\mathcal{T}_{\mu} \lessapprox 2$ for $\mu \in \{x, y\}$. For strong couplings, the rotor excitations localize along both lattice directions, forming a Ferroelectric Small Polaron (FSP). In this state, the rotors localize the charge carrier to a few sites, thus resulting in $\mathcal{T}_{\mu}\to 0$ for both $\mu \in \{x, y\}$.

As mentioned above, this approach based on electron localization is limited to the regime where $V_0 \gg B$. A complementary variational analysis shows that for weak couplings, $V_0\sim B$, the ground state is more accurately described by an extended, delocalized polaron ansatz analogous to the vGH ansatz presented in~\cite{Koutentakis2023}. In this ansatz, we describe the electron in a plane wave basis and formulate the rotor wavefunction in the co-moving frame of the electron, which inherently accounts for rotor-electron correlations. The variational optimization of the vGH ansatz improves the FLP ground states of the localized ansatz in a regime where $V_0 \lesssim 0.35t$ and the highly asymmetric domain-wall states for $| \Delta_t |\gtrsim 0.7$~\cite{Footnote}. In the remaining parameter regimes, the vGH ansatz gives rise to FDW$_{\mu}$ and FSP states with overestimated energy compared to the localization-based ansatz. Nevertheless, the vGH states have comparable physical characteristics in terms of $\boldsymbol{\mathcal{P}}^{jl}_{\text{rot}}$, $\mathcal{N}_{\text{ele}}^{jl}$ and $\mathcal{T}_{\mu}$.   

\begin{figure}[b]
\includegraphics[width=0.95\linewidth]{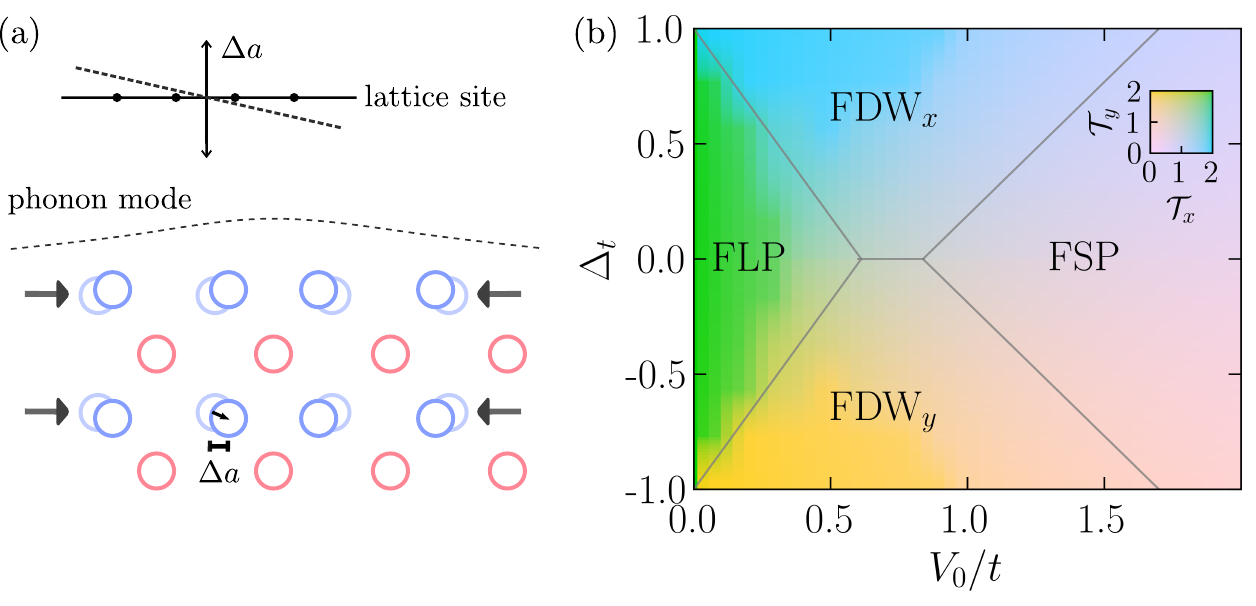}  
\vskip -0.4cm
\caption{(a) Segment of a long-wavelength, tunneling modulating lattice deformation displacing the electron sites (blue) along the $x$ axis. This deformation mode extends over at least the domain-wall region and locally results in $\Delta_t > 0$. (b) Combined variational ground state phase diagram for $B = 10^{-2}t$, considering only the state of minimum energy. The order parameters are $\mathcal{T}_{\mu}$ and the grey lines indicate the sharp phase transitions within the vGH ansatz. 
\label{fig:fig_phase_diagram}}
\end{figure}

In Fig.~\ref{fig:fig_phase_diagram}(b), we show the phase diagram in the $\Delta_t$--$V_0$ plane resulting from both variational approaches, considering only the state of minimum energy~\cite{Footnote}. For symmetric lattices, $\Delta_t = 0$, there is a smooth transition from the FLP to the FSP states as one goes from the weak to the strong coupling regime. The introduction of a finite tunneling anisotropy, $\Delta_t \neq 0$, leads to FDW$_{\mu}$ states. In the limiting case $\Delta_t\rightarrow\pm 1$, the FDW$_\mu$ states are the ground states even for weak couplings, $V_0 \to 0$,~cf.~\cite{Koutentakis2023}.

\begin{figure}[h]
\includegraphics[width=\linewidth]{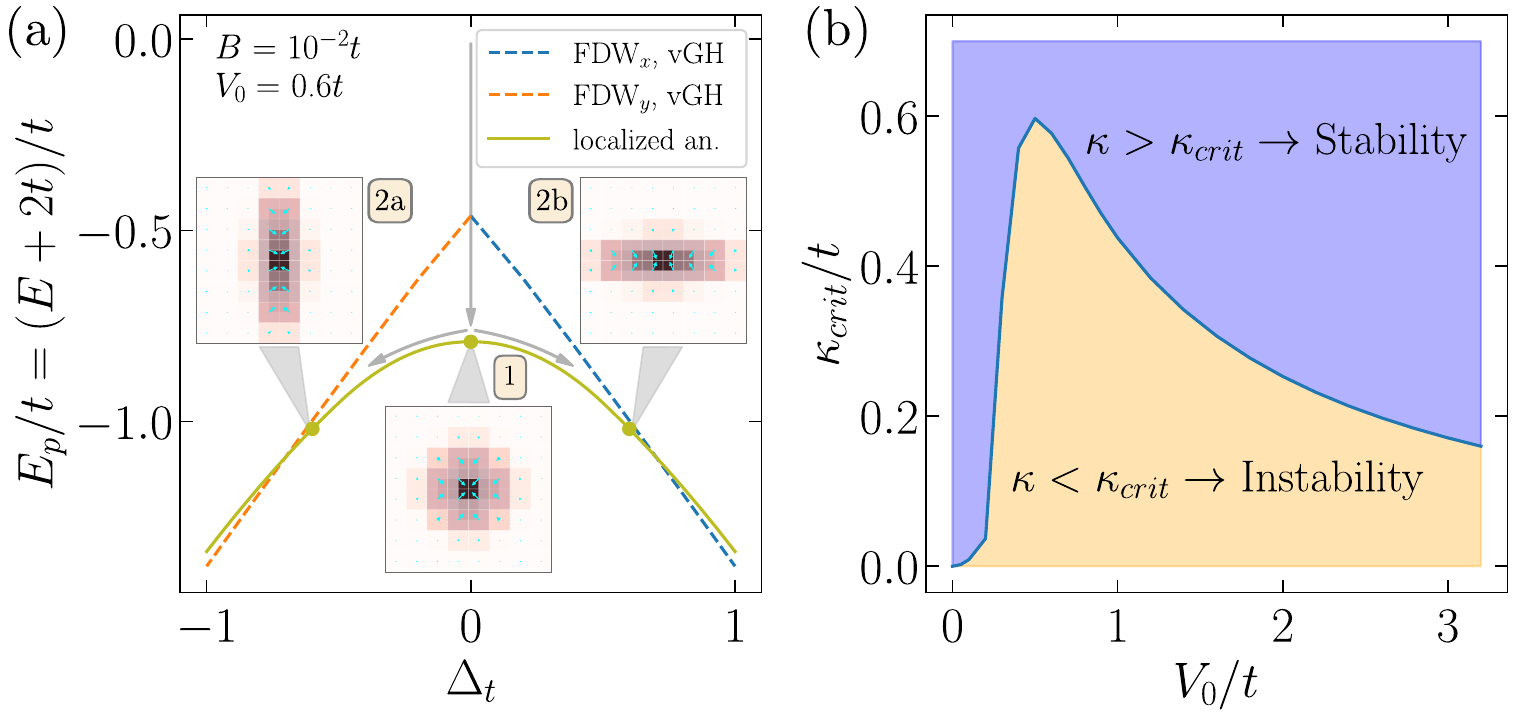}  
\vskip -0.4cm
\caption{(a) Variational ground state energy dependence on the tunneling anisotropy for an intermediate coupling, $V_0 = 0.6t$. The two-step process leading to the formation of domain-wall polarons is illustrated by the grey arrows and yellow boxes. (b) Phase diagram for a stability-instability transition, based on the critical lattice stiffness  as a function of the coupling strength for $B=10^{-2}t$.
\label{fig:fig_instability}}
\end{figure}

The dependence of the energy on $\Delta_t$ shows a significant energy benefit for increasing anisotropy, see Fig.~\ref{fig:fig_instability}(a), indicating an instability towards anisotropic states. To quantitatively estimate the regime of this instability, we expand the electronic Hamiltonian $\hat{H}_{\text{hop}}$ in terms of long-wavelength tunneling modulating lattice displacements up to second order around $\Delta_t = 0$, assuming that the vibrational Hamiltonian favors an isotropic state. That is, we can write $\hat{H}_{\text{hop}}\rightarrow \hat{H}_{\text{hop}} + \kappa \Delta_t^2$. The parameter $\kappa$ is inversely proportional to the electron-phonon coupling and proportional to the elastic constant of the lattice~\cite{Footnote,Riera2006,AlbertoLamas2009,Zhang2020}. Based on this, the instability sets in for systems with strong electron-phonon couplings or soft lattices, i.e. where $\kappa < \kappa_{\text{crit}} = \tfrac{1}{2}\partial E^2/\partial \Delta_t^2|_{\Delta_t = 0}$. This is consistent with the notion that ferroelectric domain-wall polarons occur in materials with soft lattices~\cite{Wang2021,Wang2022}. The stability regions in terms of  $\kappa_{\text{crit}}$, see Fig.~\ref{fig:fig_instability}(b), depend on material parameters and thus make a qualitative prediction for the stability-instability transition. As expected, the system is largely stable in the weak coupling limit, $V_0\rightarrow 0$, as the polaron is in a delocalized state. For strong couplings, the system is also stable, as the electron becomes almost completely localized in a single lattice site. In both cases, the polaron energy depends very weakly on $\Delta_t$. For intermediate couplings, however, $\kappa_{\text{crit}}$ peaks around $V_0\sim 0.5 t$, indicating the presence of a structural instability as the polaron energy strongly decreases for increasing anisotropy, see Fig.~\ref{fig:fig_instability}(a) and ~\cite{Footnote}.

This instability is a manifestation of the pseudo-Jahn-Teller symmetry-breaking mechanism~\cite{Bersuker2006}, as quantum fluctuations associated with the electron-phonon interaction can reduce the total energy of the system due to the net energetic benefit of the $\Delta t \neq 0$ configuration provided by FDW$_{\mu}$ formation. Within the vGH ansatz, the FDW$_{\mu}$ states are degenerate for $\Delta_t = 0$, which can be lifted for $\Delta_t \neq 0$, see Fig.~\ref{fig:fig_instability}(a). Therefore, Jahn-Teller symmetry breaking~\cite{englman1972jahn,Teller1937,Bersuker2006} is expected, independent of the stability properties of the electron-rotor system. In the localized ansatz, the degeneracy at~$\Delta_t = 0$ is lifted by correctly accounting for the coupling between electron and rotors. This leads to a cutoff in terms of $\kappa$ to trigger $C_4$ to $C_2$ symmetry reduction modes, and hence, the system exhibits the pseudo-Jahn-Teller effect~\cite{Bersuker2006}. These modes can be provided by tunneling modulating electron-phonon interactions, see Fig.~\ref{fig:fig_phase_diagram}(a), if the corresponding lattice stiffness parameter is weak enough~\cite{Footnote}.

As a result of this mechanism, the ferroelectric domain-wall polaron formation is a two-step process. First, the electron localizes, forming FLP states. Then, the energy of the system relaxes further through structural distortions of the inorganic sublattice, giving rise to a tunneling anisotropy and symmetry broken FDW$_{\mu}$ state. This two-step mechanism was previously proposed for FDW$_{\mu}$ formation in LHPs based on DFT case studies~\cite{Ambrosio2018,Ambrosio2019,Wang2021}. Furthermore, the instability condition obtained within our model, $V_0\sim t$, agrees with typical parameters in LHPs~\cite{Fabini2017,Kang2017}.

In conclusion, we have studied the phase diagram of an effective, material-independent, two-dimensional rotor-lattice polaron model with two variational approaches. In the regime of small rotational constants, there are four distinct phases, each characterized by the collective ordering of the dipole moments. Based on energetic considerations, we put forth a two-step polaron formation process, which involves charge localization in the dipolar field of the rotors, followed by pseudo-Jahn-Teller symmetry-breaking structural distortions. The resulting anisotropic wavefunction connects our model to the ``Belgian waffle'' polarons theoretically proposed in soft semiconductors \cite{Wang2021,Wang2022}.

To rigorously map these ferroelectric polarons to those observed in soft semiconductors, further investigations should identify the origin of the effective dipoles, study the transport characteristics and analyze the three-dimensional generalization where the rotors have even richer possibilities to control the tunneling direction of the electron. Finally, experimentally testing our prediction regarding structural instability could identify new materials that host domain-wall ferroelectric polarons.

\begin{acknowledgments}
We thank, in alphabetical order, Zhanybek Alpichshev, Cesare Franchini, Areg Ghazaryan, Sebastian Maehrlein and Artem Volosniev for fruitful discussions and comments. G.M.K. received funding from the European Union’s Horizon 2020 research and innovation program under the Marie Skłodowska-Curie grant agreement No.~ 101034413. M.L.~acknowledges support by the European Research Council (ERC) Starting Grant No.~801770 (ANGULON).
\end{acknowledgments}

\bibliography{bibliography}

\end{document}